\documentclass[12pt,preprint]{aastex}

\input psfig.sty

\slugcomment{To appear in ApJL}

\shorttitle{Eccentric giant planets and terrestrial planet formation}
\shortauthors{Veras \& Armitage}

\begin{document}

\title{Influence of massive planet scattering on nascent terrestrial planets}

\author{Dimitri Veras\altaffilmark{1,2} and Philip J. Armitage\altaffilmark{1,2}}
\altaffiltext{1}{JILA, Campus Box 440, University of Colorado, Boulder CO 80309; 
Dimitri.Veras@Colorado.EDU; pja@jilau1.colorado.edu}
\altaffiltext{2}{Department of Astrophysical and Planetary Sciences, University of Colorado, Boulder CO 80309}

\clearpage
\pagebreak
\newpage

\begin{abstract}
In most extrasolar planetary systems, the present orbits of known giant planets admit the existence of 
stable terrestrial planets. Those same giant planets, however, have typically 
eccentric orbits that hint at violent early dynamics less benign for low mass planet formation.
Under the assumption that massive planet eccentricities are the end point of gravitational scattering 
in multiple planet systems, we study the evolution of the building blocks of terrestrial planets during 
the scattering process. We find that typically, evolutionary sequences that result in a moderately 
eccentric giant planet orbiting at $a \simeq 2.5 \ {\rm AU}$ eject over 95\% of the material initially 
present within the habitable zone. Crossing orbits largely trigger the ejection, and leave the 
surviving material with a wide dispersion in semi-major axis, 
eccentricity and inclination. Based on these results, we predict that radial velocity follow-up of 
terrestrial planet systems found by {\em Kepler} will find that these are anti-correlated with the 
presence of eccentric giant planets orbiting at a few AU.
\end{abstract}


\section{Introduction}
Precision radial velocity surveys of nearby FGKM stars show that at least 6\% 
of them harbor giant planets \citep{marcy04}, with that fraction rising 
rapidly for metal-rich stars \citep{santisramayo01,fischer04}. In the most popular models  
for planet formation \citep{pollack96,chambers04}, the same metal-rich conditions 
which hasten the growth of giant planet cores also   
promote the formation of terrestrial planets at smaller orbital radii. Naively, 
therefore, one expects that stars hosting giant planets ought to be 
promising locations for finding terrestrial planets, although the 
converse need not be true \citep{kokubo02}. This conclusion is not 
immediately challenged by the existence of short-period and eccentric 
giant extrasolar planets, since orbit integrations have shown that the orbital 
radii and eccentricities of giant extrasolar planets do not normally preclude the 
existence of stable terrestrial planets 
\citep{levilissdunc98, jonesleecham01,dvorsuli02,noblmusicunt02,menou03,asghetal04}. Indeed, the 
majority of systems allow dynamically stable habitable terrestrial planets 
as defined by \cite{kasting93}, although formation calculations show that 
the final masses and orbital radii of terrestrial planets are likely to 
vary depending on the configuration of the giant planets \citep{leviagno03}.
In some circumstances, terrestrial planets---or the building blocks thereof---
may even survive giant planet migration \citep{mandsigu03}.

In this letter, we investigate whether the {\em early} evolution of multiple 
giant planets in extrasolar planetary systems could be more restrictive 
of terrestrial planet formation than the relatively benign dynamical 
environment that exists today. The typical 
extrasolar planet with a semi-major axis of a few AU has an 
eccentric orbit \citep{marcy04}, most likely as a consequence 
of either planet-disk interactions \citep{artymowicz91,papaloizou01,goldreich03} 
or gravitational scattering in multiple planet systems 
\citep{rasio96,lin97,levilissdunc98,ford01,marzari02}. Although both 
mechanisms are theoretically plausible, to date only the scattering hypothesis has been 
demonstrated to provide a quantitative match to the distribution 
of extrasolar planet eccentricities \citep{ford03}. Here, we assume 
that scattering is the origin of the eccentricity of extrasolar 
planets, and study the influence of such scattering on orbits within the terrestrial
planet zone (a related study was performed by Thommes, Duncan \& Levison 1999, who discussed the
impact of outward migration of Uranus and Neptune on the Kuiper Belt).  The initial
conditions for our scattering experiments are described in \S 2.  In \S 3, we describe
our main result - that the early dynamical evolution expected within an unstable
multiple planet system acts to clear out a large fraction of the mass that would
otherwise be available to form terrestrial planets at smaller radius.  We suggest
in \S 4 that this effect might be strong enough to produce an anti-correlation between
the presence - in the same system - of eccentric massive planets at a few AU and
terrestrial planets.

\section{Initial conditions}

For our initial conditions, we assume that several massive planets have 
formed from the protoplanetary disk in a configuration that is stable 
over short time scales (of the order of the gas dissipation time scale, 
which is approximately $10^5$~yr), but unstable over longer time scales. 
Depending upon exactly how multiple giant planets typically form, such 
a gas-free interim state may or may not exist. Ignoring not only any 
residual gas, but also dynamical friction from planetesimal swarms
\citep{wethstew93}, we study the coupled evolution of the massive 
planets together with a population of terrestrial planet cores at 
smaller orbital radii.  We justify our neglect of dynamical friction by
noting that the Jovian planets are substantially more massive
than the surviving solid bodies in the vicinity.  At the time when the gas disk is dissipated, 
several tens of such cores---which for our purposes can be treated 
as massless test particles---are likely to exist (Kominami \& Ida 2002, 2004) and
be concentrated in the habitable zone \citep{wetherill96}.

Eccentric massive planets can result from dynamical evolution in two planet 
\citep{rasio96}, three planet \citep{marzari02}, or initially more crowded 
systems \citep{lin97,levilissdunc98,papaterq01,adams03}. Which of these (if any) is physically 
most appropriate as an initial condition is unknown. Following \cite{marzari02}, 
we adopt an initial state with three giant planets, as this is the simplest 
system that yields instability on the desired time scales without considerable 
fine tuning of the initial planetary separations. 

Using Rauch \& Hamilton's (2004, in preparation) HNbody Bulirsch-Stoer integrator, which 
is designed for the integration of N-body systems with a single massive object, we 
performed $7$ integrations.
Each integration modeled the evolution of $3$ Jovian-mass planets and
$50$ test particles in the terrestrial planet region around 
a solar-mass star over $2$ Myr. The Jovian-mass planets acted on the terrestrial ones, but not vice versa.
In each simulation, all planets were started on circular, almost coplanar orbits, with randomly assigned 
angular variables and randomly assigned inclination values less than $0.001^{\circ}$.
At time intervals of $10^4$ yr, if the periapse of a test particle 
was smaller than
one Solar radii, we considered that test particle to have collided with the star.
The semimajor axes of the test particles were evenly distributed from $0.75$ AU to
$1.25$ AU.  For the massive planets, we fixed $a_2 = 6$~AU and $a_3 = 12$~AU, 
and varied $a_1$ in order to sample the range of possible behavior. For six cases 
$j = 1-6$, we set $a_1 = 4.80 + 0.02(j-1)$~AU. Case $7$ is a repeat
of case $6$ except with different random choices of the angular variables and inclinations. 
With this setup, in all but one case, energy
and angular momentum errors, expressed by $(E - E_0)/E_0$ and $(L - L_0)/L_0$, where
$E_0$ and $L_0$ are the initial system energy and angular momentum, did not exceed $10^{-9}$.

The motivation for our choice of initial giant planet separations was the desire to study systems 
which have dynamically settled into a state resembling known multi-planet exosystem configurations 
- configurations featuring two well-separated eccentric giant planets.  By targeting the final location
of such an eccentric giant planet in the semimajor axis range, $2-3$ AU, we found through preliminary
tests that the initial semimajor axis of the innermost planet should exceed $4.5$ AU.  In order to produce the
instability needed to eject one giant planet on timescales exceeding several orbits but less than several 
million years, we set the ratio $a_2/a_1$ between 1.20 and 1.22, just within the globally chaotic limit 
established by \cite{wisdom80} for two-planet systems.  Our choice of $a_3/a_2 = 2$ was arbitrary,
but did prove through experimentation to yield the types of systems we sought to investigate.

\section{Results}

The final orbital parameters of the giant planets in each of the seven systems simulated 
are summarized in Table 1. Asterisks denote a giant planet that was ejected from the system; 
in 5 of the 7 simulations, one planet was ejected. The final states of these 5 systems are 
broadly similar---one  planet settled into an orbit with a semimajor
axis of between $2$ AU - $3$ AU, while the other achieved an orbit with $a \ge 9$ AU. Figure 1
illustrates the evolution of the orbital parameters for the three planets in case $1$. One 
observes that the outer planet
interacts significantly with the middle planet after $\sim 1$ Myr, is then shoved outward beyond
$50$ AU at about $1.45$ Myr, and finally is ejected from the system right before $2$ Myr has elapsed.
The middle planet settles into an orbit of $a_2 \approx 9$ AU and the inner planet settles with
$a_1 \approx 3$ AU.  From Table 1, one sees that the middle planet's orbit was nearly circularized
after $2$ Myr, while the inner planet, whose semimajor axis varied by no more than a few AU 
throughout the simulation, attained a significant eccentricity of over $0.2$. The evolution seen here 
is representative of the typically complex interactions that occur in a multiple planet 
system prior to ejection, although it cannot be emphasized too much that the system is 
chaotic and that several qualitatively different outcomes are possible.  Further, the systems 
in Table 1 do not appear to have settled down; more dynamical evolution is likely, some of 
it violent.  A comparison of case 6 and case 7 illustrates explicitly how different outcomes can arise 
even though the same initial semimajor axis and eccentricity values were used in both cases.

Figure 2 displays the fraction of the 50 test particles,
all initially residing within a loosely defined habitable zone of $0.75-1.25$ AU, that survived
the $2$ Myr-long shuffling of the three giant planets.  Triangles show the fraction of test 
particles that survived {\it and} still reside in the habitable zone; diamonds the total fraction 
that survived with semimajor axes $< 100$ AU.  The solid line is the average value 
(1\%) of the triangles, and the dotted line is the average value (14\%) of the diamonds. 
All of the test particles that survive outside the habitable zone are propelled 
outward beyond $1.25$ AU rather than inward within $0.75$ AU. Overall, almost all
(94\% - 100\%) of the
terrestrial material were cleared out of the habitable zone. In cases 1, 4, 5, and 6, not a single
test particle survived in the habitable zone. If the simulations continued for longer
than $2$ Myr, the remaining test particles would likely dwindle to zero.

The study of \cite{leviagno03} shows that where terrestrial planets form is likely 
to depend on the configuration of giant planets within the same system. Systems 
in which there is significant dynamical excitation of planetary embryos around 1~AU 
form a smaller number of planets, which lie closer to the star. Massive planet 
scattering provides an extreme example of such dynamical excitation.  In hot and 
dynamically excited enough systems, collisions
between growing embryos are, in fact, more likely erosive then accretional. 
Although, as shown in Table 1, all Jovian-mass planets remain in prograde orbits with 
inclinations less than $52^{\circ}$, the inclinations of the test particles are 
scattered over the entire $0^{\circ}-180^{\circ}$ range, even for those particles that remain in 
the habitable zone.  Retrograde test particles orbits are triggered by crossing
orbits with the prograde Jovian planets.  Further, the eccentricities of the test 
particles that remain in the
habitable zone after $2$ Myr are highly variable and encompass the entire allowable range. 
This distribution suggests that even in those systems where enough material survives in 
the inner planetary system to form terrestrial planets, the characteristics of the resulting 
planets could differ between systems where scattering of giant planets has 
occurred and those where it has not. By analogy with the calculations of \cite{leviagno03}, 
one might expect that terrestrial planets that form subsequent to giant planet 
scattering would typically lie closer to the star, and possibly have significant 
inclinations relative to the surviving giant planets. Such planets would be a 
stark contrast to what is seen in the Solar System.

By design, the final state of the systems that we have simulated includes an 
innermost giant planet with orbital properties similar to those of many known 
extrasolar planetary systems (significant eccentricity, semi-major axis of a 
few~AU). If we consider {\em only} this innermost planet, thereby ignoring 
both the possibility of additional planets today and the dynamical history, 
then the habitable zone admits relatively stable circular orbits within 
which terrestrial planets might form. Specifically, as shown in Table~2, 
if we integrate case 1 forward an additional 2~Myr with only the innermost 
giant planet and a new population of test particles on circular orbits in 
the habitable zone, then nearly a quarter survive. This result conforms well to that 
of \cite{menou03} for real extrasolar planetary systems. We also investigated 
the stability of orbits in the inner planetary system of case 1 when the 
existence of the outer planet---which survived the initial evolution leading 
to ejection of one planet---was considered. In this case interactions 
between the two massive planets lead to variations in their eccentricity; these
variations slowly eject test particles placed (again) on initially 
circular orbits at smaller radii.  Reasonably plausible planet formation 
scenarios that result in eccentric giant planets are thus doubly 
hazardous to terrestrial planet formation---not only because of the initial 
round of ejections but also because of subsequent secular evolution which can sweep 
clear much of the material in the inner planetary system.

\section{Discussion}

The main result of this letter is that terrestrial planet formation is 
likely to be more difficult in planetary systems in which there is an 
eccentric giant planet at moderate orbital radius (2 to 3~AU for the 
specific configuration studied here), {\em if} giant planet eccentricity 
is a consequence of scattering in young multiple planet systems.
This conclusion implies that - at fixed stellar metallicity - there may be fewer
terrestrial planets orbiting stars with massive planets in this semimajor
axis range than around stars with more distant giant planets.  However, an 
observed lack of terrestrial planets in such systems does not necessarily 
imply that they were molded by gravitational instability. 
The existence of a possible anti-correlation between the presence of eccentric 
giant planets and terrestrial planets should be testable using the sample of 
terrestrial planets discovered by the {\em Kepler} transit mission \citep{borucki03}. 
With plausible assumptions, {\em Kepler} 
should identify of the order of $10^2$ stars with terrestrial planets, along 
with around 30 giant planets with $a > 1.6 \ {\rm AU}$. Given even modest 
relative inclinations between giant and terrestrial planets in the same system, 
the small yield of outer orbit giants means that {\em Kepler} on its own will 
not provide a statistical picture of the typical configuration of extrasolar 
planetary systems. Spectroscopic follow-up of the stars harboring terrestrial planets, 
however, will provide useful information on the presence or absence of outer 
giant planets. {\em Kepler} target stars for the terrestrial planet search 
have $m_V < 15$, and should be bright enough to allow radial velocity 
measurements with precision of tens of meters per second (for comparison, 
existing observations of OGLE-TR-56 by \cite{konacki03} attain a precision 
of $\approx 100 \ {\rm ms^{-1}}$ at $m_V = 16.6$). Radial velocity surveys 
show that about 2\% of target stars have massive planets in the range 
$2 \ {\rm AU} < a < 4 \ {\rm AU}$, with that fraction rising rapidly 
for metal-rich stars \citep{fischer04}. There should, therefore, be 
enough terrestrial planets found for radial velocity follow-up to 
either confirm or rule out the anti-correlation with eccentric 
giant planets proposed in this paper.

We thank Hal Levison for his insights and suggestions as well as for 
volunteering to compare our results to those from his own code.  We gratefully 
acknowledge support from the National Science Foundation under
grant AST~0407040, and from NASA under grant NAG5-13207 issued through
the Office of Space Science.

\pagebreak

\pagebreak

\begin{figure}
\plotone{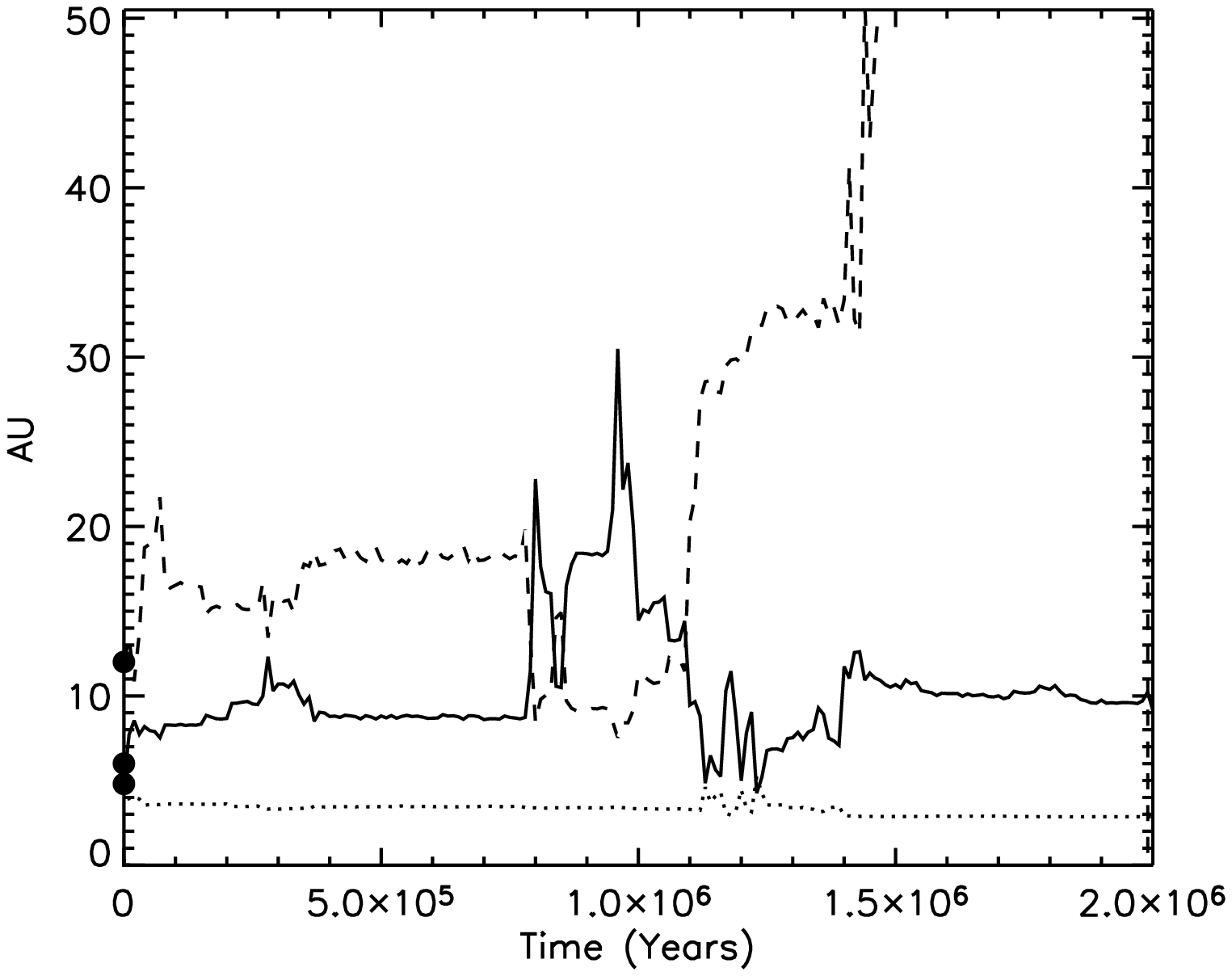} 
\label{case6}
\figcaption{Orbital evolution over $2$ Myr of all three Jovian-mass planets in case $1$.  
Thick dots indicate initial locations of planets.  The evolution of the initially inner, 
middle, and outer planets are illustrated through the dotted, solid, and dashed lines
respectively.}
\end{figure}

\pagebreak

\begin{figure}
\plotone{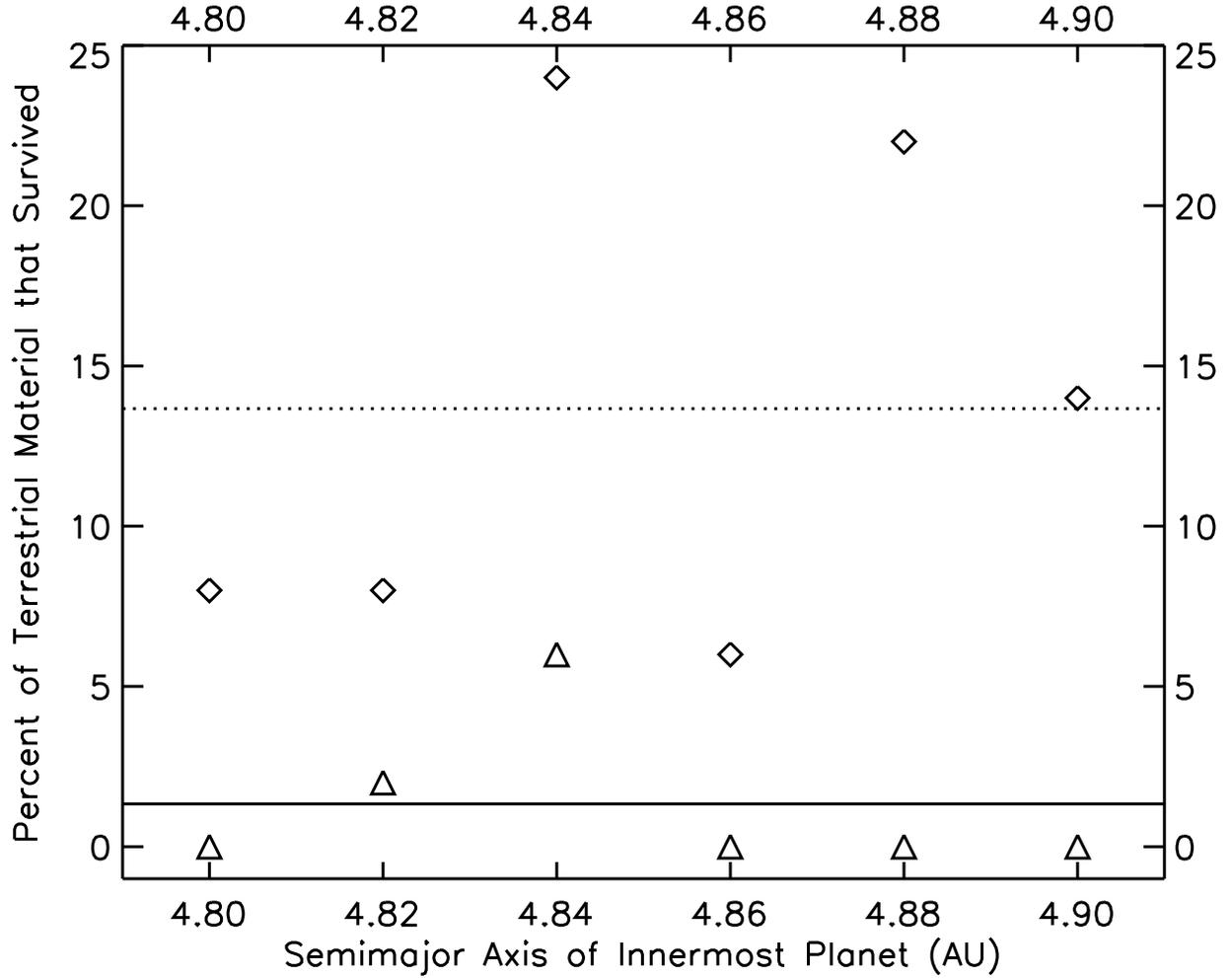} 
\label{masrem}
\figcaption{Percent of terrestrial planets which survive over $2$ Myr.  The
horizontal axis contains the initial semimajor axis values of the innermost giant planet
in cases 1-6, but otherwise has no physical significance.  Triangles indicate
planets whose final positions reside in the habitable zone ($0.75-1.25$ AU).  Diamonds
indicate planets whose final semimajor axes are within $100$ AU.  The horizontal solid
and dotted lines represent the average value of the triangles and diamonds, obtained
from cases $1 - 6$.}
\end{figure}

\pagebreak

\begin{deluxetable}{cccccccccc}
\tabletypesize{\small}
\tablecaption{GIANT PLANET OUTCOMES\label{plan}}
\tablecolumns{10}
\tablewidth{0pt}
\tablehead{\colhead{case} & 
   \colhead{$a_1$} &
   \colhead{$a_2$} & 
   \colhead{$a_3$} &
   \colhead{$e_1$} & 
   \colhead{$e_2$} & 
   \colhead{$e_3$} &
   \colhead{$I_1$} &
   \colhead{$I_2$} &
   \colhead{$I_3$} 
}
\startdata
1 & 9.1 & 2.9 & $\ast$ & 0.23 & 0.025 & $\ast$ & 9.5 & 40 & $\ast$ \\
2 & $\ast$ & 2.4 & 28 & $\ast$ & 0.082 & 0.38 & $\ast$ & 34 & 2.3 \\
3 & 4.4 & 64 & 4.7 & 0.19 & 0.83 & 0.31 & 26 & 12 & 26 \\
4 & 8.6 & 2.0 & 150 & 0.31 & 0.28 & 0.94 & 15 & 31 & 19 \\
5 & $\ast$ & 2.3 & 74 & $\ast$ & 0.77 & 0.42 & $\ast$ & 52 & 21 \\
6 & 2.3 & 37 & $\ast$ & 0.46 & 0.54 & $\ast$ & 8.6 & 8.5 & $\ast$ \\
7 & 39 & $\ast$ & 2.3 & 0.68 & $\ast$ & 0.13 & 3.9 & $\ast$ & 12 \\
\enddata
\tablecomments{Final orbital parameters for the three planets (innermost denoted
by ``1'', outermost denoted by ``3'') in all seven cases, each run for $2$ Myr.
Semimajor axes and inclinations are in units of AU and degrees respectively.
Asterisks indicate planets that were ejected from the system.}
\end{deluxetable}

\clearpage
\pagebreak

\begin{deluxetable}{ccc}
\tabletypesize{\small}
\tablecaption{CASE 1 EXTENSIONS - SURVIVAL PERCENTAGES\label{plan2}}
\tablecolumns{3}
\tablewidth{0pt}
\tablehead{\colhead{Time (Myr)} & 
   \colhead{2 Giant Planets} & 
   \colhead{1 Giant Planet}
}
\startdata
0 & 100 & 100 \\
0.2 & 62 & 78 \\
0.4 & 46 & 74 \\
0.6 & 32 & 68 \\
0.8 & 28 & 58 \\
1.0 & 18 & 44 \\
1.2 & 14 & 34 \\
1.4 & 12 & 32 \\
1.6 &  8 & 30 \\
1.8 &  8 & 28 \\
2.0 &  4 & 24 
\enddata
\tablecomments{Percent of test particles which reside in the habitable zone as a function of
time and initial conditions for the two $2$ Myr extensions performed on case 1.  The 
designations ``2 Giant Planets'' and ``1 Giant Planet'' reflect these two simulations, described in 
the text. 
}
\end{deluxetable}

\end{document}